\documentclass[useAMS,usenatbib]{mn2e}

\usepackage{graphicx}

\title[Stellar metallicity and orbital periods]
{On the possible correlation between the orbital periods 
of extra-solar planets and the metallicity of 
the host stars}
\author[A. Sozzetti]{A. Sozzetti$^{1,2,3}$\thanks{E-mail:
alex@phyast.pitt.edu}\\
$^{1}$University of Pittsburgh, Dept. of Physics \&
Astronomy, Pittsburgh, PA 15260, USA \\
$^{2}$Harvard-Smithsonian Center for Astrophysics, 
60 Garden Street, Cambridge, MA 02138, USA \\
$^{3}$INAF - Osservatorio Astronomico di Torino, 10025 Pino
Torinese, Italy}
\begin{document}

\date{Accepted ????. Received ???}

\pagerange{\pageref{firstpage}--\pageref{lastpage}} \pubyear{2002}

\maketitle

\label{firstpage}

\begin{abstract}
We investigate a possible correlation between the orbital periods 
$P$ of the extra-solar planet sample and the metallicity [Fe/H] 
of their parent stars. Close-in planets, on a few-days orbits, 
are more likely to be found around metal-rich stars. 
Simulations show that a weak correlation is present. 
This correlation becomes stronger when only sigle stars 
with one detected planet are considered. We discuss several potential sources 
of bias that might mimic the correlation, and find they can be ruled out, 
but not with high significance. If real, the absence of 
very short-period planets around the stellar sample with [Fe/H] $< 0.0$ 
can be interpreted as evidence of a metallicity dependence of the 
migration rates of giant planets during formation in the protoplanetary 
disc. The observed $P-$[Fe/H] correlation can be falsified or confirmed 
by conducting spectroscopic or astrometric surveys of metal-poor stars 
([Fe/H] $< -0.5$) in the field.
\end{abstract}


\section{Introduction}

The nearby F-G-K dwarfs harbouring giant planets 
show evidence of moderate metal-enrichment with respect to the average 
metallicity of field dwarfs in the solar neighborhood. The 
dependence of planetary frequency on the metallicity of 
the host stars was investigated since the first detections by precision 
radial-velocity surveys (Gonzalez 1997; Laughlin \& Adams 1997), 
and different explanations were 
proposed, such as enhanced giant planet formation by high stellar 
metallicity (Santos et al. 2000, 2001; Reid
2002), observational selection effects or pollution by ingested planetary 
material (Laughlin 2000; Gonzalez et 
al.\ 2001; Israelian et al. 2001; Pinsonneault et al. 2001; Murray \& Chaboyer 2002). 
Recently, based on observationally unbiased stellar samples, the evidence for 
higher planetary frequency around unpolluted, primordially metal-rich 
stars has been clearly demonstrated by Santos et al. (2001), and confirmed by 
Fischer et al. (2003) and by Santos et al. (2004), 
who showed a sharp break in frequency at [Fe/H] $\simeq 0.0$. In this 
paper we investigate further the possible correlation between the orbital 
period of the extra-solar planet sample and the metallicity of 
the parent stars, in favor of which some authors had argued in the 
past (Gonzalez 1998; Queloz et al. 2000; Jones 2003), 
while others (Santos et al. 2001; Laws et al  2003) had not found 
evidence of its existence. 

In a recent work, Santos et al. (2003) discussed extensively 
possible dependencies 
between stellar and planetary properties. In particular, they concluded 
that the {\it metallicity} distribution of stars with very short-period 
planets ($P \leq 10$ days) is essentially indistinguishable from the 
same distribution of stars with longer period ($P > 10$ days) planets. 
In this paper we show instead that there appears to be some evidence 
for a significant difference between the {\it period} distributions of planets 
around metal-rich ([Fe/H] $ \geq 0.0$) and metal-poor ([Fe/H] $ < 0.0$) 
stars. This correlation between stellar metallicity and orbital 
periods is highlighted by a paucity of close-in planets ({\it Hot Jupiters} 
on circular orbits with $P \leq 5$ days) around the metal-poor stellar 
sample. 

In Section 2 we present our statistical studies of the $P-$[Fe/H] 
correlation for the extra-solar planet sample. In Section 3 we analyze 
possible sources of bias that might contribute to produce the 
observed trend. In Section 4 we briefly present our findings in the 
context of formation and, in particular, migration processes for giant 
planets in primordial protoplanetary discs, and discuss possible 
observational tests that might help disprove or verify the reality 
of the correlation. 

\section[]{Stellar and Planet Sample Analysis}

\begin{table*}
\caption{Metallicities of planet-host stars and orbital periods of the 
planetary-mass companions utilized in the analysis. 
The list is sorted by increasing period of the 
innermost planet. The literature source used for the metallicity values 
is Santos et al. (2004), except for HD 330075 (Pepe et al. 2004), 
BD-10 3166 (Gonzalez et al. 2001), and HD 37605 (Cochran et al. 
2004).}

\label{tab1}
\begin{minipage}{\textwidth}
\centering
\begin{tabular}{@{}lcccc}
\hline
Star  & [Fe/H] & 
\multicolumn{3}{c}{Orbital Period (days)
\footnote{Data from http://www.obspm.fr/encycl/encycl.html as of July 2004}}\\
& & Planet 1 & Planet 2 & Planet 3 \\
\hline
 HD 73256   &   0.26 & 2.548 & & \\
 HD 83443   &   0.35 & 2.985 & & \\
 HD 46375   &   0.20 & 3.024 & & \\
 HD 179949  &   0.22 & 3.093 & & \\
 HD 187123  &   0.13 & 3.097 & & \\
$\tau$ Boo\footnote{Star in a binary system 
(Eggenberger et al. 2004, and references therein)}& 0.23 & 3.313 & & \\
 HD 330075  &   0.08 & 3.369 & & \\
 BD -10 3166 &   0.33 & 3.487 & & \\
 HD 75289\footnote{Star in a binary system 
(Mugrauer et al. 2004a)} & 0.28 & 3.510 & & \\
 HD 209458  &   0.02 & 3.525 & & \\
 HD 76700   &   0.41 & 3.971 & & \\
 51 Peg     &   0.20 & 4.231 & & \\
 $\upsilon$ And$^b$   &   0.13 & 4.617 &241.5 &1284.0 \\
 HD 49674   &   0.33 & 4.948 & & \\
 HD 68988   &   0.36 & 6.276 & & \\
 HD 168746  &  -0.08 & 6.403 & & \\
 HD 217107  &   0.37 & 7.110 & & \\
 HD 162020  &  -0.04 & 8.428 & & \\
 HD 130322  &   0.03 & 10.72 & & \\
 HD 108147  &   0.20 & 10.90 & & \\
 HD 38529  &   0.40 & 14.31 &2174.3 & \\
 55 Cnc$^b$    &   0.33 & 14.65 &44.28 &5360.0  \\
 GJ 86$^b$      &  -0.24 & 15.78 & & \\
 HD 195019$^b$  &   0.08 & 18.30 & & \\
 HD 6434    &  -0.52 & 22.09 & & \\
 HD 192263  &  -0.20 & 24.35 & & \\
 $\varrho$ Crb    &  -0.21 & 39.85 & & \\
 HD 74156  &   0.16 & 51.64 & 2300.0 & \\
 HD 37605  &   0.39 & 54.23 & & \\
 HD 168443 &   0.06 & 58.12 &1739.5 & \\
 HD 3651    &   0.12 & 62.23 & & \\
 HD 121504  &   0.16 & 64.60 & & \\
 HD 178911$^b$  &   0.27 & 71.49 & & \\
 HD 16141   &   0.15 & 75.56 & & \\
 HD 114762$^b$  &  -0.70 & 84.03  & & \\
 HD 80606$^b$   &   0.32 & 111.8 & &\\
 70 Vir     &  -0.06 & 116.7 & & \\
 HD 216770  &   0.26 & 118.5 & & \\
 HD 52265   &   0.23 & 119.0 & & \\
 HD 1237    &   0.12 & 133.8 & & \\
 HD 37124  &  -0.38 & 152.4 &1495.0 & \\
 HD 73526   &   0.27 & 190.5 & & \\
 HD 82943  &   0.30 & 221.6 &444.6 & \\
 HD 169830 &   0.21 & 225.6 &2102.0 & \\
 HD 8574    &   0.06 & 228.8 & &\\
 HD 89744\footnote{Star in a binary system 
 (Wilson et al 2001; Mugrauer et al. 2004b)} & 0.22 & 256.6 & &  \\
 HD 134987  &   0.30 & 260.0 & & \\
 HD 12661  &   0.36 & 263.6 & 1444.5 & \\
 HD 150706  &  -0.01 & 264.9 & & \\
 HD 40979$^b$   &   0.21 & 267.2 & & \\
 HD 17051   &   0.25 & 320.1 & & \\
 HD 142     &   0.14 & 338.0 & & \\
 HD 92788   &   0.32 & 340.0 & & \\
 HD 28185   &   0.22 & 385.0 & & \\

\hline
\end{tabular}
\end{minipage}
\end{table*}

\begin{table*}
\contcaption{}
\begin{minipage}{\textwidth}
\centering
\begin{tabular}{@{}lccccc}
\hline
Star & [Fe/H] & 
\multicolumn{3}{c}{Orbital Period (days)$^a$}\\
& & Planet 1 & Planet 2 & Planet 3 \\
\hline
 HD 142415  &   0.21 & 386.3 & & \\
 HD 177830  &   0.33 & 391.0 & & \\
 HD 108874  &   0.23 & 401.0 & & \\
 HD 4203    &   0.40 & 401.0 & & \\
 HD 128311  &   0.03 & 414.0 & &\\
 HD 27442   &   0.39 & 423.8 & &\\
 HD 210277  &   0.19 & 437.0 & & \\
 HD 19994$^b$   &   0.24 & 454.0 & & \\
 HD 20367   &   0.17 & 500.0 & & \\
 HD 114783  &   0.09 & 501.0 & & \\
 HD 147513  &   0.06 & 540.4 & & \\
 HD 222582  &   0.05 & 572.0 & & \\
 HD 65216   &  -0.12 & 613.1 & & \\
 HD 160691  &   0.32 & 638.0 &1300.0 & \\
 HD 141937  &   0.10 & 653.2 & & \\
 HD 23079   &  -0.11 & 738.5 & & \\
 16 Cyg B$^b$    &   0.08 & 799.0 & & \\
 HD 4208    &  -0.24 & 812.2 & & \\
 HD 114386  &  -0.08 & 872.0 & & \\
 HD 213240  &   0.17 & 951.0 & & \\
 HD 10647   &  -0.03 &1040.0 & &\\
 HD 10697   &   0.14 &1078.0 & & \\
 47 UMa    &   0.06 &1095.0 &2594.0 & \\
 HD 190228  &  -0.26 &1127.0 & & \\
 HD 114729  &  -0.25 &1131.5 & & \\
 HD 111232  &  -0.36 &1143.0 & & \\
 HD 2039    &   0.32 &1192.6 & & \\
 HD 136118  &  -0.04 &1209.6 & & \\
 HD 50554   &   0.01 &1279.0 & &\\
 HD 196050  &   0.22 &1289.0 & & \\
 HD 216437  &   0.25 &1294.0 & & \\
 $\tau^1$ Gruis  &   0.24 &1443.0 & & \\
 HD 106252  &  -0.01 &1500.0 & & \\
 HD 23596   &   0.31 &1558.0 & & \\
 14 Her     &   0.43 &1796.4 & & \\
 HD 39091   &   0.10 &2063.8 & & \\
 HD 72659   &   0.03 &2185.0 & & \\
 HD 70642   &   0.18 &2231.0 & & \\
 HD 33636   &  -0.08 &2447.3 & & \\
 $\varepsilon$ Eri    &  -0.13 &2502.1 & &  \\
 HD 30177   &   0.39 &2819.7 & &  \\
 GJ 777 A$^b$    &   0.24 &2902.0 & & \\
\hline
\end{tabular}
\end{minipage}
\end{table*}

\begin{figure*}
$\begin{array}{cc}
\includegraphics[width=0.48\textwidth]{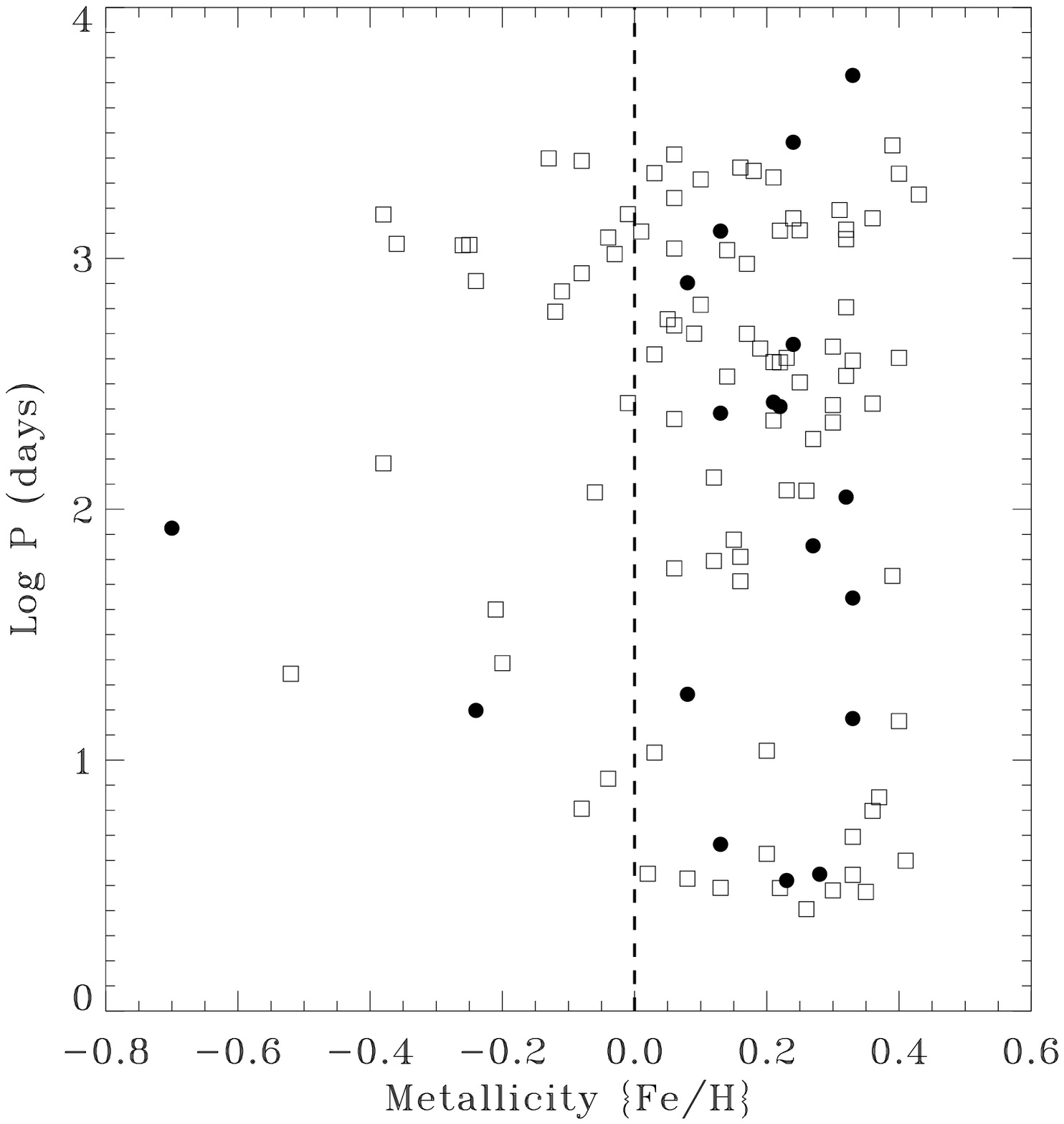} & 
\includegraphics[width=0.48\textwidth]{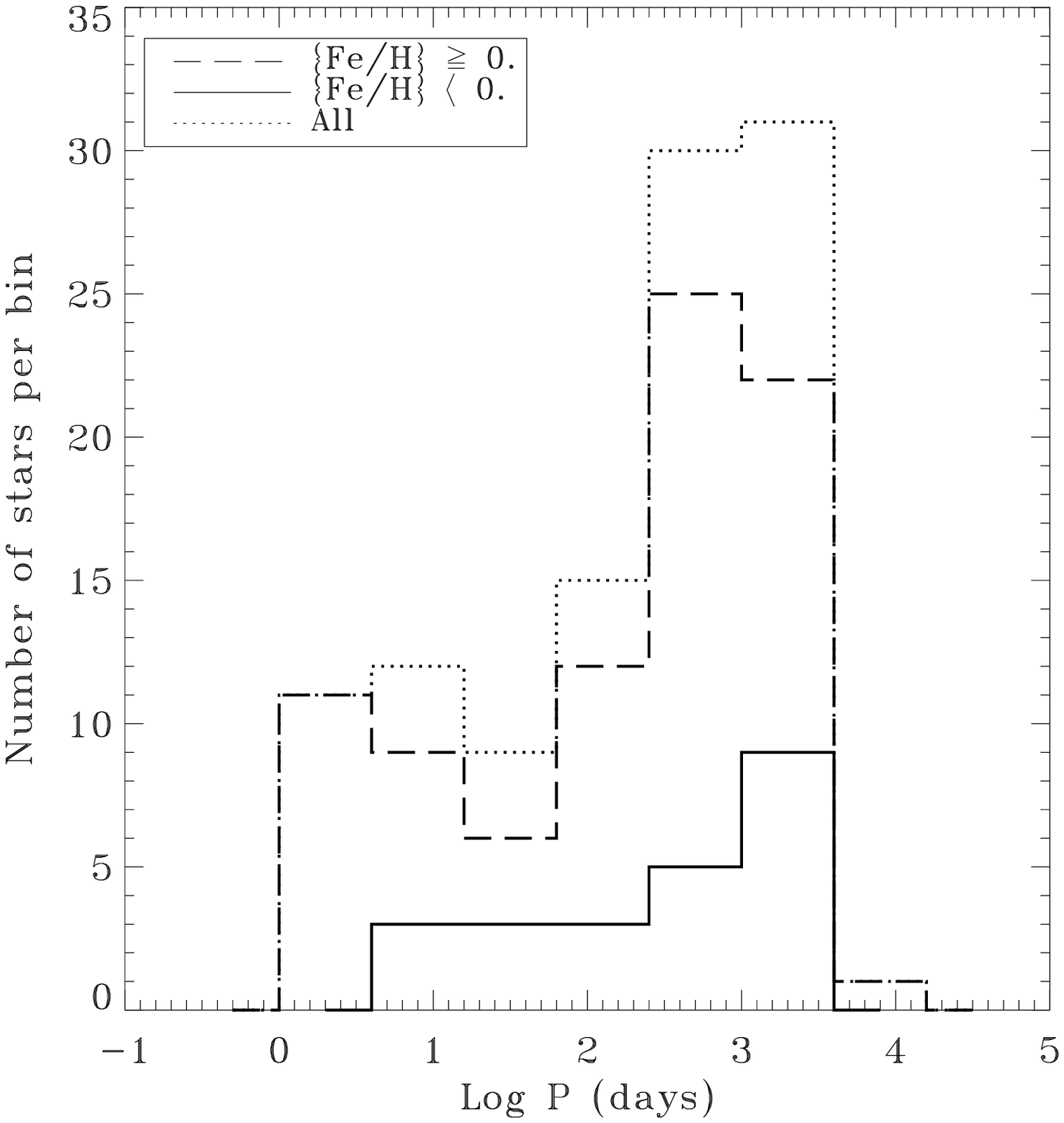}
\end{array} $
\caption{Left panel: orbital periods of extra-solar planets 
as a function of the metallicity of the host stars. Planets identified by filled circles 
are orbiting known members of binary systems. Right panel: distribution of 
orbital periods for the stellar sample with [Fe/H] $< 0.0$ (solid line), with 
[Fe/H] $\geq 0.0$ (dashed line), and for the full sample (dotted line)}
\label{fig1}
\end{figure*}

We summarize in Table 1 the values of orbital periods $P$ for the sample of 
extra-solar planets known to-date and the metallicities [Fe/H] of their parent stars 
that have been utilized in our analysis. As described in the Table, 
orbital periods were taken from up-to-date online catalogues, while the 
metallicity values were collected from a variety of sources, primarily a 
recent paper by Santos et al. (2004). The detailed 
list of literature sources used is reported in Table 1, along with the relevant 
information about binarity. This sample of 96 stars, and 109 planets, is the 
result of the adoption of a few selection criteria we believe are important in 
order to keep the possible sources of bias at a minimum. The impact of additional 
potential biases, that cannot be removed by simply excluding a few objects from 
the analysis, will be discussed in Section 3. For the purpose of our study, 
we have excluded from the sample the following objects:

\begin{enumerate}
\item [1)] sub-stellar companions to nearby stars with minimum projected masses in the 
brown dwarf mass regime. Given the still uncertain and evolving definition 
of a giant planet, the dividing line must be set with some degree of 
arbitrariness, and it may ultimately turn out that brown dwarfs and 
planets indeed populate a common mass range and/or a common origin. However, 
in our case we have used a more relaxed version of the 
Oppenheimer et al. (2000) theoretical Deuterium-burning threshold of 
13 M$_\mathrm{J}$ (where M$_\mathrm{J}$ is the mass of Jupiter), 
that establishes both the 
lower limit to the mass of a brown dwarf and the upper bound to the
mass of a planet (assuming solar metallicity). 
In particular, we have excluded objects with masses exceeding 
this limit by more than 25-30\%, 
except for the case of the multiple system orbiting HD 168443, which probably 
shares a common origin;
\item [2)] six spectral class III K-G giants (HD 219449, HD 104985, HD 59686, Hip 75458, 
HD 47536, and $\gamma$ Cep), belonging to different samples of stars with 
respect to the original F-G-K class IV-V subgiant/dwarf stars included in 
the observing lists of the major precision Doppler surveys for planets;
\item[3)] the unconfirmed second planet around $\varepsilon$ Eri;
\item[4)] the first recently discovered planetary mass object OGLE-235/MOA-53 by 
means of the microlensing technique (Bond et al. 2004), as the 
characteristics of the parent star are not well determined;
\item[5)] the two strongly interacting planets orbiting the M4 dwarf GJ 876 and 
the planet around HD 41004 A, 
for which no reliable metallicity estimates  have been provided yet;
\item[6)] the three recently announced ``very'' Hot Jupiters orbiting OGLE 
transiting candidates (e.g., Konacki et al. 2003; Bouchy et al. 2004). Indeed, 
for two of these objects (OGLE-TR-113 and OGLE-TR-132) Bouchy et al. (2004) 
have provided metallicity estimates, but the low S/N ratios of the spectra for these 
stars do not allow at present such parameter to be well constrained. Furthermore, the 
metallicity distribution of the OGLE sample (at a typical distance of about 
1.5 kpc) is unknown, and it might significantly 
differ from the one of the solar neighborhood sample (within 40-50 pc of the Sun) 
observed by current precision radial-velocity surveys. In a recent work, 
Nordstr\"om et al. (2004) confirmed the existence of a mild radial 
metallicity gradient in the disc of the Milky Way, and its evolution with time. 
In particular, stars younger than $\sim 10$ Gyr show an average metallicity 
gradient of $\sim -0.09$ dex/kpc, while the oldest stars in their sample do not 
show any gradient at all. On the face of it, it is then safer to exclude the OGLE 
transiting planets from the sample.
\end{enumerate}

In Figure 1, left panel, we show the log-distribution 
of $P$ as a function of [Fe/H). According to Santos et al. (2004), the percentage 
of planet host stars increases linearly with [Fe/H] for metallicity 
values greater than solar, while it flattens out for metallicities 
lower than solar. We then divide the orbital period distribution 
into two metallicity bins ([Fe/H] $< 0.0$ and [Fe/H] $\geq 0.0$), 
and compare them in the histogram plot in the right panel of 
Figure 1. For reference, the full distribution of orbital 
periods for all metallicities is also overplotted. The most striking 
feature arising from the plot is the total absence of close-in 
planets on $P\leq 5$ days, circularized orbits around the metal-poor stellar 
sample. We must then ask if this effect is statistically significant. 
In order to do so, several tests can be conducted. We opt for a 
Kolmogorov-Smirnov (K-S) test, to measure to what extent the two period 
distributions might differ, and for a rank correlation test, to 
verify whether the period and metallicity distributions are actually 
uncorrelated. 
As for the latter, we prefer to evaluate the (Spearman or Kendall) 
rank-order correlation coefficient rather than the classic linear 
(Pearson) correlation coefficient. In fact, non-parametric, rank correlation 
analyses are on average more robust, and the significance of a negative or 
positive rank-order correlation coefficient can usually be assessed 
(see for example Press et al. 1992). 

\begin{table}
\centering
\caption{Results of the K-S and rank correlation tests on different 
stellar subsets: the full sample (1), when binaries are removed 
(2), when stars with multiple-planet systems are removed (3), and 
when both binaries and stars with planetary systems are removed. For 
completeness, we also list, for each case studied, 
the numbers $N_\mathrm{rich}$ and $N_\mathrm{poor}$ of planets orbiting 
stars in the two metallicity bins ([Fe/H] $\geq 0.0$ and [Fe/H] $< 0.0$, 
respectively).}
\label{tab2}
\begin{tabular}{@{}ccccc}
\hline
Sub-sample & $Pr(D)$ & $Pr(r_s)$ & $N_{\mathrm{rich}}$ & $N_{\mathrm{poor}}$\\
\hline
1 & 0.09 & 0.11 & 86 & 23 \\
2 & 0.09 & 0.11 & 70 & 21 \\
3 & 0.06 & 0.04 & 64 & 21 \\
4 & 0.03 & 0.01 & 54 & 19 \\
\hline
\end{tabular}
\end{table}
The computation 
of the relevant $D$ statistics in the K-S test gave as a result $D = 0.26$, 
corresponding to a probability of the two period distributions to be 
the same $Pr(D) \simeq 0.09$. A measure of the relevant statistics $r_s$ 
for the rank-order (Spearman) correlation test gave $r_s = -0.14$, with 
a corresponding probability of the two distributions to be uncorrelated 
$Pr(r_s) \simeq 0.11$. There is then evidence for marginal differences 
in the period distributions for metal-poor and metal-rich stars, as 
well as a weak anti-correlation between the $P$ and [Fe/H] distributions. 

To further investigate the possible existence of such effect, 
we have performed the same two 
tests on three sub-samples of the full dataset, i.e., removing stars in 
binary or multiple stellar systems, stars with multiple-planet systems, 
and both, respectively. The results are summarized in Table~2. 
As it can be seen, the K-S test is not very sensitive to different sub-samples 
of stars with planets, while the correlation between the $P$ and [Fe/H] 
distributions gets more significant, especially when only single stars 
orbited by a single planet are taken into account. Anyhow, the two tests are in 
fair agreement with each other. The lack of sensitivity 
of the K-S test is possibly due to an intrinsic property of the test itself. 
In fact, the K-S test is most sensitive around the median value of a given 
cumulative distribution function, and less sensitive at the extreme ends of the 
distribution. For this reason the test probably works best in detecting 
shifts in a probability distribution, which will likely affect its median value, 
while it might fail to detect spreads, that would primarily affect the tails of 
a probability distribution rather than its median. 
This is quite likely our case, as the main difference between the 
distributions of planet orbital periods for the metal-poor and metal-rich sample 
resides in the very short-period regime, i.e. at one of the tails of the distribution.
Note, however, that any hint of correlation disappears if one removes from the analysis 
the stars harbouring the Hot Jupiters with $P\leq 5$ days. For comparison, 
analogous K-S and rank correlation tests were performed on the planet 
mass distribution split in two metallicity bins, and probabilities of the null 
hypotheses to be the correct ones in the ranges $0.22\leq Pr(D)\leq 0.44$ 
and $0.38\leq Pr(r_s)\leq 0.87$ were obtained, 
respectively. In this case, similarly to the findings of Santos et al. 
(2003), no significant trend was revealed.

In order to further quantify the statistical significance of the 
results on the $P-$[Fe/H] correlation, we have utilized Monte Carlo 
simulations, in a fashion similar to the analyses undertaken 
by Zucker \& Mazeh (2002) and Mazeh \& Zucker (2003) in their works 
on the mass-period correlation of extra-solar planets and the possible 
mass ratio-period ratio correlation in multiple-planet systems. 
We have randomly drawn 109 pairs of points (corresponding to the present 
total number of planets announced, with the constraints discussed above) 
from the {\it observed} log-distributions of orbital periods of the extra-solar 
planet sample and metallicities of the host stars, and repeated the 
process $10^5$ times, calculating the rank correlation coefficient 
for each simulated dataset. 

\begin{figure}
\includegraphics[width=0.45\textwidth]{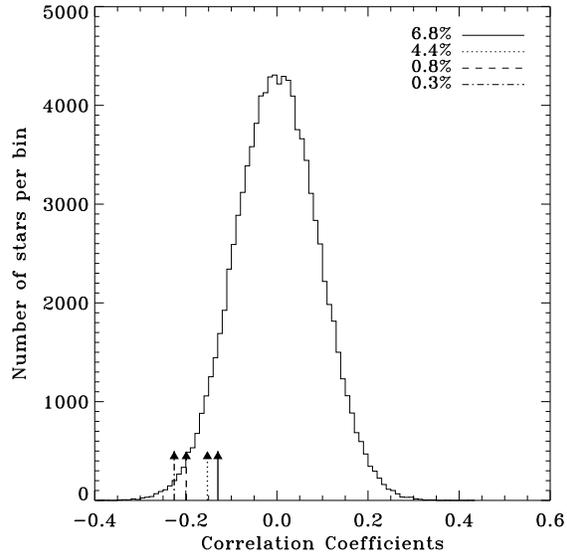}
\caption{Histogram of the rank correlation coefficients calculated from $10^5$ random 
samples drawn from the observed stellar metallicity and orbital period 
distributions. The observed values discussed in Section 2 (arrows of different 
styles) have between 6.8\% and 0.3\% chances of being consistent with no 
correlation}
\label{fig2}
\end{figure}

In Figure 2 we show the histogram 
of the simulated values of the $r_s$ statistics, together with the 
observed values we obtained for the full stellar sample and the three 
sub-sets defined above. As a general result, both the intrinsic probability 
estimates of the statistical tests and the simulations we performed agree 
in indicating that the null hypothesis that 
orbital period and metallicity distributions are uncorrelated is rejected 
with a confidence level in the range 93.2\%-99.7\%, corresponding to a 
2- to 3-$\sigma$ result. Indeed, the evidence for a correlation is somewhat 
weak, but the pile-up of very close-in planets preferably around 
metal-rich parent stars, if real, may have important consequences for 
our understanding of some crucial aspects of the formation and migration 
processes of gas giant planets.

\section{Observational Biases?}

There are at least three main sources of potential observational biases in 
the data (see Gonzalez (2003) for a thorough review of the subject): 
1) there could be significant errors in the determination of 
the metal content of the host stars, 
2) the sample of extra-solar planets around stars with [Fe/H] $< 0.0$, 
only 1/3 of the number of planets detected around 
metal-rich field dwarfs, might not be large enough for 
statistical analyses; and 3) metal-poor stars have weaker spectral lines 
with respect to solar-type dwarfs, so that they are in principle more 
difficult targets for high-precision radial-velocity surveys. 

In the first case, all the metallicity values we utilized (see Table 1) have 
been derived by means of spectroscopic analysis methods of high-resolution, 
high-S/N echelle spectra. As pointed out in recent works (e.g. Laws et al. (2003); 
Gonzalez (2003), and references therein), significant systematic offsets 
can be found between spectroscopic and photometric [Fe/H] determinations, 
and ultimately spectroscopic methods seem to be more reliable. 
With this approach, the typical uncertainties reported in 
abundance analyses for planet-host stars are of order of a few 
hundredths of a dex. Within these limits, then according to Santos 
et al. (2004) 12 of the 14 Hot Jupiters with $P\leq 5$ days included in our 
sample would still orbit metal-rich stars, and the same holds for HD 330075 
and BD -10 3166, according to Pepe et al. (2004) and Gonzalez et al. (2001), 
respectively. In conclusion, uncertainties in the [Fe/H] determination can be ruled out, 
at least to first approximation, as possible causes of the observed correlation between 
extra-solar planets' orbital periods and metallicities of the parent 
stars. However, if the metallicities of a couple of the stars harbouring 
Hot Jupiters were ill-determined due for example to some unrecognized systematics 
and they turned out to be falling in the range $-0.2\leq \mathrm{[Fe/H]}\leq -0.1$, 
then this would significantly dilute the effect. 

Secondly, due to the lower planet occurrence 
rate around metal-poor stars, the absence of Hot Jupiters around the metal-poor 
stellar sample could indicate that simply not enough objects have been 
observed yet in that metallicity bin. 
However, comparable sample sizes of stars in the two metallicity 
bins have been monitored for several years by precision Doppler surveys 
(e.g., Fischer et al. 2003; Santos et al. 2003). There are 23 planets in the 
[Fe/H] $< 0.0$ bin, and 86 in the [Fe/H] $\geq 0.0$ bin. In the latter, 
14 objects are orbiting the parent stars with $P\leq 5$ days, about $16\%\pm 4\%$ 
of the full sample (assuming Poisson statistics). 
If the occurrence rate for Hot Jupiters in the 
low-metallicity bin is comparable, then we should expect about $4\pm 2$ 
planets to be orbiting with $P\leq 5$ days, but there are no detections 
in this period range. This is about a 2-$\sigma$ deviation. As summarized in Table 3, 
this difference is present up to periods of order of 20-25 days. Above this 
threshold, the two fractional values become the same. Furthermore if we 
divide the [Fe/H] $\geq 0.0$ sample in two bins, $0.0\leq\mathrm{[Fe/H]}\leq 0.25$ 
and [Fe/H] $> 0.25$, what is observed is that, for example, the occurrence rate of 
Hot Jupiters with $P\leq 5$ days increases by about 60\% 
from the first to the second high-metallicity 
bin: there are in fact 7 such planets out of 52 ($\sim 13\%\pm 5\%$) orbiting stars with 
$0.0\leq\mathrm{[Fe/H]}\leq 0.25$, while 7 out of 34 planets ($\sim 21\%\pm 8\%$) are 
found on $P\leq 5$ days orbits around stars in the metallicity range 
[Fe/H] $> 0.25$. However, due to the limited amount of data, uncertainties are 
large enough that such trend is even less significant than the one 
observed by comparing the [Fe/H]$< 0.0$ and the [Fe/H]$\geq 0.0$ bins. 

In the end, these results are further suggestive of a higher likelihood of finding
giant planets on close-in orbits around increasingly more metal-rich stars. 
The small-number statistics argument can thus be ruled out as a major 
contributor to the observed $P-$[Fe/H] correlation, but only at the 2-$\sigma$ 
confidence level.

\begin{table}
\centering
\caption{Fraction of planets below a given value of orbital 
period $P$ in the two metallicity bin [Fe/H] $< 0.0$ and [Fe/H] $\geq 0.0$.}
\label{tab3}
\begin{tabular}{@{}ccc}
\hline
Period interval & [Fe/H] $< 0.0$ & [Fe/H] $\geq 0.0$\\
\hline
$P\leq 5$ days & $0\%$ & $16\%\pm 4\%$ \\
$P\leq 10$ days & $9\%\pm 6\%$ & $19\%\pm 5\%$ \\
$P\leq 15$ days & $9\%\pm 6\%$ & $23\%\pm 5\%$ \\
$P\leq 20$ days & $13\%\pm 7\%$ & $24\%\pm 5\%$ \\
$P\leq 25$ days & $22\%\pm 8\%$ & $24\%\pm 5\%$  \\
\hline
\end{tabular}
\end{table}

Finally, due to the weak spectral lines, the low-metallicity objects might 
have been monitored by Doppler surveys with somewhat lower velocity 
precision, thus a fraction of the planets might have gone undetected. 
Santos et al. (2003) and Fischer et al. (2003) have studied this possibility 
by calculating 
the median velocity error as a function of metallicity for the stars in their 
planet surveys, and found a velocity degradation of up to 50\% for the lowest 
metallicity stars ([Fe/H]$\simeq -0.5$). Radial-velocity surveys currently 
attain typical single-measurement precisions $\sigma_{\mathrm{RV}}\simeq 3-5$ 
m/s, and given the fact 
that the most glaring discrepancy between the orbital period distributions for 
planets around low- and high-metallicity is the absence of close-in planets, 
which would be easily detected (we recall the radial velocity amplitude 
$K\propto P^{-1/3}$) even with  $\sigma_{\mathrm{RV}}\simeq 5-8$ m/s, 
then we can conclude that also radial-velocity precision degradation 
for metal-poor stars is not 
a major cause for the observed correlation (in the long period regime 
the datasets typically contain observations with lower precision, say 10-15 m/s, 
but in this limit the two period distributions do not present any differences). 
However, the fine details of the observing strategies for the two stellar 
samples are not known exactly, and there is a non-zero chance that 
some bias might be introduced by human factors (e.g., less observing 
time spent on the metal-poor sample, more on the metal-rich sample with a 
higher chance of planet discovery announcements).

\section{Discussion}

We have presented new intriguing evidence for a lack of planets on 
very short-period orbits ($P\leq 5$ days) around stars with metallicity 
[Fe/H] $< 0.0$, confirming early findings by Gonzalez (1998) and Queloz et 
al. (2000), and more recently by Jones (2003). 
As shown through statistical tests as well as Monte Carlo 
simulations, the $P-$[Fe/H]] correlation is moderately significant (2- to 
3-$\sigma$ level), and it gets stronger when only single stars orbited by 
single planets are considered. We have discussed a variety of possible 
sources of observational biases, and did not find any strong evidence of 
them playing a significant role in the determination of the observed 
correlation. However, potential biases introduced by uncertainties in the 
determination of the metallicities of the planet-host stars and the 
small-number statistics cannot be ruled out with very high confidence, thus 
a clear conclusion is difficult to draw at this point. 

On the other hand, if this trend is real, then the paucity of short-period giant 
planets around metal-poor stars should be explained in principle within 
the scenarios of their formation and in the context of the migration processes 
protoplanets are likely to undergo while embedded in the primordial 
protoplanetary disc. For the purpose of this analysis, we focus on the `cleaner' 
sample of single stars orbited by single planets, which exhibits the stronger 
$P-$[Fe/H] correlation, as in presence of multiple planets and/or binary stellar systems 
the outcome of formation and/or migration could be significantly different 
(e.g., Zucker \& Mazeh (2002), and references therein; Mazeh \& Zucker 
(2003), and references therein; Eggenberger et al. (2004), and references therein). 
This is however subject to change if for example a significant fraction of the present-day 
single-planet systems turned up to have additional long-period companions. 

It is probably 
premature at this stage to make meaningful statements on the relative roles of the two 
proposed mechanisms for gas giant planet formation, i.e. core accretion 
(e.g., Lissauer 1993; Pollack et al. 1996; Alibert et al. 2004) 
and disc instability (e.g., Boss 1997, 2001; Mayer et al. 2002; Rice et al. 2003). 
As there is no reason to believe that 
orbital parameters distributions of giant planets formed in different ways 
around different stellar populations would be very similar (Boss 2002), one should 
in principle be able to find fossil evidence of the formation processes in such 
distributions. For example, the two formation mechanisms would predict 
quite distinct mass distributions (e.g., Ida \& Lin 2004; Rice et al. 2003), 
and the dependence of planetary frequency 
on the metallicity of the protoplanetary disc is also expected to be rather 
different (e.g., Pollack et al. 1996; Boss 2002). However, either because of a lack of 
sensitivity of present-day detection techniques at the low-mass end 
or in light of incompleteness of the different stellar populations targeted, 
no general conclusions can be drawn at present (except that maybe both mechanisms 
operate).

On the other hand, regardless of how giant planets formed, a significant fraction 
of them must have undergone some degree of orbital migration, in particular all the 
Hot Jupiters. Thus the observed period and 
eccentricity distributions of extra-solar planets, and correlations among planet 
orbital parameters and masses, are somewhat more likely to reflect 
migration-related effects, blurring the evidence in such distributions for different 
formation scenarios. Indeed, Udry et al. (2003), for example, show that the highly 
non-Gaussian distribution of orbital periods is likely to be the outcome of 
the (Type II) migration process, and the variety of mechanisms that might trigger it, 
in fair agreement with theoretical predictions (see for example Armitage et al. 
(2002) and references therein). 
Similar conclusions are also reached by Zucker \& Mazeh (2002) to justify the evidence 
for a shortage of high-mass planets in short-period orbits. 

The absence of planetary objects with $P\leq 5$ 
days around stars with [Fe/H] $< 0.0$ can also be explained within the context of migration 
scenarios. Again, we concentrate on models that do not consider interactions with a 
distant companion star or dynamical instabilities in multiple-planet systems, 
which might be required for explaining at most $\sim 20-25\%$ of the systems discovered 
so far (out of 108 stars with planets, 16 have a certified binary companion, 12 
harbour more than one giant planet, with two of the planetary systems found orbiting one 
of the components of wide binary stellar systems). In the more widely accepted 
model of (Type II) migration in a gaseous disc, a giant planet massive enough to 
open a gap around itself will become locked to the disc and will ultimately share 
its fate (e.g., Goldreich \& Tremaine 1979; Papaloizou \& Lin 1984; 
Ward 1997; Trilling et al. 2002). 
As shown by Livio \& Pringle (2003), if disc opacity $\kappa$ increases with 
increasing metallicity, if disc temperature $T$ increases with increased opacity, 
then this leads to a higher kinematic viscosity $\nu$, and this shortens the viscous 
inflow timescale $\tau_\nu$, i.e. it makes the disc evolve faster. In their work, Livio \& 
Pringle (2003) assume a weak dependence of migration timescales on metallicity 
($\tau_\nu \propto \nu^{-1} \propto 
T^{-1} \propto \kappa^{-0.34} \propto \mathrm{[Fe/H]}^{-0.34}$), 
and conclude that this effect cannot account for the observed decrease in the 
probability of a star having giant planets in the observed range of periods 
as its metallicity decreases, thus a lower occurrence rate at low values of [Fe/H] 
is indicative of lower probability of formation, not migration. However, given 
the uncertainties on some of the parameters describing the detailed structure 
of a protoplanetary disc and its evolution, and their relative dependencies, 
it is not inconceivable to argue for a more substantial dependence of migration 
rates on metallicity. 
Such explanation would fit the observed trend we are beginning to unveil, 
i.e. the much lower occurrence rate of Hot Jupiters around the metal-poor sample 
of stars with planets. Indeed, although not yet statistically significant, 
this trend seems to be 
present already in the metal-rich sample, with the fraction of Hot Jupiters 
decreasing by $\sim 60\%$ when we move from the [Fe/H] $> 0.25$ to the 
$0.0\leq \mathrm{[Fe/H]}\leq 0.25$ bin. 

In conclusion, there are several ways to improve on our understanding of the 
complex interplay between the observed properties of extra-solar planets 
(due to formation and/or migration processes) and those of the host stars. 

For what concerns the possible $P-$[Fe/H] correlation, a solid theoretical 
basis for its existence or absence (as pointed out by Livio \& Pringle (2003)) 
could be established if detailed high-resolution, three-dimensional, time-dependent 
migration computations were to be carried out, which would include a sophisticated 
treatment of the thermal structure of disc. 

On the observational side, improvements in quantitative spectroscopic analyses 
due for example to better input physics (stellar atmosphere models), high-quality 
instrumentation, and more refined measurement and analysis software may help to further 
reduce the uncertainties on metallicity determination, hopefully also for 
stars significantly cooler or hotter, as well as more active, than our Sun. 
Such efforts will nevertheless have to be coupled to an enlargement of the sample-size 
of stars with planets. 
In particular, one of the most effective ways to prove or falsify the $P-$[Fe/H] 
correlation and its potential consequences 
for orbital migration and/or giant planet formation scenarios discussed in this work 
would be to extend the sample size of the metal-poor population, including 
a statistically significant number of very metal poor ([Fe/H] $\leq -0.5$) 
objects. This could be achieved by combining radial-velocity (e.g., 
Sozzetti et al. 2003a) and high-precision astrometric searches with 
ground-based as well as space-borne observatories that will come on-line 
during the next decade or so (e.g., Sozzetti et al. 2001, 2002, 2003b). 

\section*{Acknowledgments}

We are grateful to A. P. Boss, D. W. Latham, D. D. Sasselov, and G. Torres for stimulating 
discussion and helpful insights. 
This work has greatly benefited from the comments they provided on an 
earlier version of this paper. We thank an anonymous referee for careful 
reading of the manuscript and for very important 
suggestions and advise. A. S. acknowledges support from the Smithsonian 
Astrophysical Observatory through the SAO Predoctoral Fellowship program.

\label{lastpage}

\end{document}